%% file: main_arxiv.tex
\documentclass{article}

\usepackage{arxiv}

\usepackage[utf8]{inputenc} % allow utf-8 input
\usepackage[T1]{fontenc}    % use 8-bit T1 fonts
\usepackage{hyperref}       % hyperlinks
\usepackage{url}            % simple URL typesetting
\usepackage{booktabs}       % professional-quality tables
\usepackage{amsfonts}       % blackboard math symbols
\usepackage{amsmath} 
\usepackage{nicefrac}       % compact symbols for 1/2, etc.
\usepackage{microtype}      % microtypography
\usepackage{graphicx}
\usepackage[sort&compress,comma,super]{natbib}
\usepackage{doi}

\usepackage[textsize=tiny]{todonotes}
\usepackage{graphicx}% Include figure files
\usepackage{dcolumn}% Align table columns on decimal point
\usepackage{bm}% bold math
\usepackage{hyperref}% add hypertext capabilities
\usepackage[mathlines]{lineno}% Enable numbering of text and display math
\usepackage{color}
\usepackage{cases}
\usepackage{siunitx}

\title{An efficient numerical algorithm for\\ the moment neural activation}

\author{ 
\href{https://orcid.org/0000-0001-6174-1660}{\includegraphics[scale=0.06]{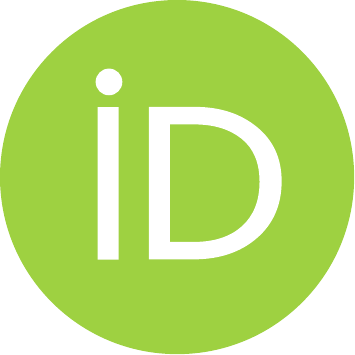}\hspace{1mm}
Yang Qi} %\thanks{Use footnote for providing further information about author (webpage, alternative address)---\emph{not} for acknowledging funding agencies.} 
\\
	Institute of Science and Technology for Brain-Inspired Intelligence \\
	Fudan University\\
	Shanghai, China\\
	\texttt{yang\_qi@fudan.edu.cn}
}

\renewcommand{\headeright}{}
\renewcommand{\undertitle}{}
\renewcommand{\shorttitle}{An efficient numerical algorithm for the moment neural activation}

\hypersetup{
pdftitle={An efficient numerical algorithm for the moment neural activation},
pdfauthor={Yang Qi},
pdfkeywords={},
}

\begin{document}
\maketitle

\begin{abstract}
Derived from spiking neuron models via the diffusion approximation, the moment activation (MA) faithfully captures the nonlinear coupling of correlated neural variability. 
However, numerical evaluation of the MA faces significant challenges due to a number of ill-conditioned Dawson-like functions. By deriving asymptotic expansions of these functions, we develop an efficient numerical algorithm for evaluating the MA and its derivatives ensuring reliability, speed, and accuracy. We also provide exact analytical expressions for the MA in the weak fluctuation limit. Powered by this efficient algorithm, the MA may serve as an effective tool for investigating the dynamics of correlated neural variability in large-scale spiking neural circuits.
\end{abstract}

% keywords can be removed

%start main text here

\input{algorithms/intro_ma}

\input{algorithms/moment_activation}

\input{algorithms/tmp_recycled}

\input{algorithms/fast_dawson}
\input{algorithms/fast_maf}
\input{algorithms/intnfire}
\input{algorithms/discus}

%appendix
\input{algorithms/derivatives}

\section*{Acknowledgments}
Supported by MOE Frontiers Center for Brain Science, Fudan University, Shanghai, China.
%\end{acknowledgments}

\section*{Code Availability}
The code for the proposed algorithm is available at: \url{https://github.com/BrainsoupFactory/moment_nn}

\bibliographystyle{abbrvnat}%plainnat}%unsrtdin}
%\bibliographystyle{unsrtnat}
%\bibliography{references}  %%% Uncomment this line and comment out the ``thebibliography'' section below to use the external .bib file (using bibtex) .

\input{reference_list}




% \end{thebibliography}

\end{document}

%% file: algorithms/intro_ma.tex
\section{Introduction}

%introduce diffusion approximation
Understanding how neurons in the brain process noisy spiking inputs with correlated fluctuations is a problem central to neuroscience\citep{Urai2022NatNeur}. A couple of mathematical techniques have been developed to analyze the statistical and dynamical properties of spiking neurons. The first of them is known as the diffusion approximation~\citep{Capocelli1971Kyber}, in which the synaptic current generated by input spikes is replaced by a Gaussian white noise with the same mean and variance. By solving the first passage time problem associated with firing threshold, the mean and variance of the output spike train can then be derived~\citep{Feng2006,Richardson07}. 
The second technique is the linear response theory which can be used to obtain the pair-wise correlation map of spiking neurons~\citep{Rocha2007nature,LU2010913}. Together, these analytical techniques lead to closed-form expressions for the mapping from the second-order statistical moments of the input synaptic current to that of the output spike train~\citep{LU2010913}. We refer to this mapping as the moment activation (MA). The MA captures the nonlinear coupling of correlated neural variability and can be thought as a natural extension of rate-based neural activation functions commonly used in modeling studies of neural circuits.

%state the challenge
Despite the availability of close-form expressions for the MA, it contains a group of ill-conditioned Dawson-like functions which render it numerically intractable~\citep{LU2010913}. First, the Dawson-like functions involve products of exploding and vanishing terms, causing their direct numerical evaluation unreliable. Second, these ill-conditioned functions occur in nested integrals, which are slow to evaluate even for input range where they are well behaved. 
Although methods such as threshold-integration schemes can be used to evaluate the MA by numerically solving the associated Fokker-Planck equation~\cite{Richardson07,Rosenbaum2016a}, these methods are computationally cumbersome and unsuitable for large population size. These challenges for numerical implementation of the MA limit its usefulness as a practical tool for analyzing the dynamics of correlated neural variability in neural circuits. 

% what this paper is about
In this study, we develop an efficient numerical algorithm for evaluating the MA ensuring both reliability and speed through a combination of techniques including asymptotic expansion and Chebyshev polynomial approximation. The proposed algorithm leads to accurate and reliable evaluation of the MA orders of magnitude faster than brute force methods. Powered by this efficient algorithm, the MA can serve as an effective tool for investigating the firing statistics and correlated variability in large-scale spiking neural circuits.

%% file: algorithms/moment_activation.tex
\section{The moment neural activation}

Consider the leaky integrate-and-fire (LIF) neuron model with membrane potential dynamics described by
\begin{equation}
\dfrac{dV_i}{dt}= -LV_i(t) + I_i(t),
%dV_i= -LV_i(t)dt + dI_i(t),
\label{eq:LIF}
\end{equation} 
where the sub-threshold membrane potential $V_i(t)$ of a neuron $i$ is driven by the synaptic current $I_i(t)$ and $L=0.05$ \si{\per\milli\second} is the conductance. When the membrane potential $V_i(t)$ exceeds a threshold $V_{\rm th}=20$ \si{\milli\volt} a spike is emitted. Afterwards, the membrane potential $V_i(t)$ is reset to the resting potential $V_{\rm res}=0$ mV, followed by a refractory period of $T_{\rm ref}= 5$ ms. Then, the mean firing rate $\mu_i$ and the firing covariabiility $C_{ij}$ can then be defined as
\begin{equation}
\mu_i = \lim_{\Delta t\to\infty} \dfrac{\mathbb{E}[N_i(\Delta t)]}{\Delta t}, 
\label{eq:def_mean}
\end{equation}
and
\begin{equation}
C_{ij} = \lim_{\Delta t\to\infty} \dfrac{{\rm Cov}[N_i(\Delta t),N_j(\Delta t)]}{\Delta t}, 
\label{eq:def_cov}
\end{equation}
where $N_i(\Delta t)$ is the spike count of neuron $i$ over a time window $\Delta t$. The moment activation (MA) essentially maps the statistical moments $(\bar{\mu}_i,\bar{C}_{ij})$ of the input current $I_i(t)$ in Eq.~\ref{eq:LIF} to the moments $(\mu_i,C_{ij})$ of output spike train~\citep{Feng2006,LU2010913}.

The first part of the MA describes the statistical input-output relation of a single neuron~\citep{Feng2006}, in which case we drop the neuronal index for clarity
\begin{numcases}{\phi: (\bar{\mu},\bar{\sigma}) \mapsto(\mu,\sigma^2),~~~~}
\mu =  \dfrac{1}{T_{\rm ref} + \mathbb{E}[T]},\label{eq:mu}
\\
\sigma^2 = \mu^3{\rm Var}[T].\label{eq:sigma}
\end{numcases}
The quantity $T$ is the inter-spike interval whose mean and variance are given by
\begin{equation}
\mathbb{E}[T] = \tfrac{2}{L}\int_{I_{\rm lb}}^{I_{\rm ub}} g(u) du = \tfrac{2}{L}[G(I_{\rm ub})-G(I_{\rm lb})],
\label{eq:mean_isi}
\end{equation}
\begin{equation}
{\rm Var}[T] = \tfrac{8}{L^2}\int_{I_{\rm lb}}^{I_{\rm ub}} h(u) du = \tfrac{8}{L^2}[H(I_{\rm ub})-H(I_{\rm lb})],
\label{eq:var_isi}
\end{equation}
with integration bounds $I_{\rm ub}(\bar{\mu},\bar{\sigma}) = \tfrac{V_{\rm th}L-\bar{\mu}}{\sqrt{L}\bar{\sigma}}$ and
$I_{\rm lb}(\bar{\mu},\bar{\sigma}) = \tfrac{V_{\rm res}L-\bar{\mu}}{\sqrt{L}\bar{\sigma}}$. The four Dawson-like functions that appear in Eq.~\ref{eq:mean_isi} and Eq.~\ref{eq:var_isi} are
\begin{equation}
g(x)=e^{x^2}\int_{-\infty}^x e^{-u^2}du,
\label{eq:g}
\end{equation}
\begin{equation}
h(x)=e^{x^2}\int_{-\infty}^x e^{-u^2}[g(u)]^2du,
\label{eq:h}
\end{equation}
\begin{equation}
G(x)=\int_0^x g(u)du
\label{eq:G}
\end{equation}
\begin{equation}
H(x)=\int_{-\infty}^x h(u)du.
\label{eq:H}
\end{equation}

The mapping for the correlation coefficients are approximated through a linear response theory as~\citep{LU2010913}
\begin{equation}
\rho_{ij}=\chi(\bar{\mu}_i,\bar{\sigma}_i)\chi(\bar{\mu}_j,\bar{\sigma}_j)\bar{\rho}_{ij},~~~~i\neq j,
\end{equation}
with 
\begin{equation}
\chi(\bar{\mu},\bar{\sigma})=\dfrac{\bar{\sigma}}{\sigma}\dfrac{\partial\mu}{\partial\bar{\mu}}.
\label{eq:linear_res}
\end{equation} 
Note that the covariance matrix $C_{ij}$ is related to the correlation coefficient $\rho_{ij}$ as $C_{ij}=\rho_{ij}\sigma_i\sigma_j$. %Note that the derivative $\partial\mu/\partial\bar{\mu}$ is evaluated as a function of $(\bar{\mu},\bar{\sigma})$. 
The linear response theory provides a first order approximation around $\bar{\rho}_{ij}=0$ by assuming that higher order terms are negligible.

\begin{figure*}
\centering
\includegraphics[width=\textwidth]{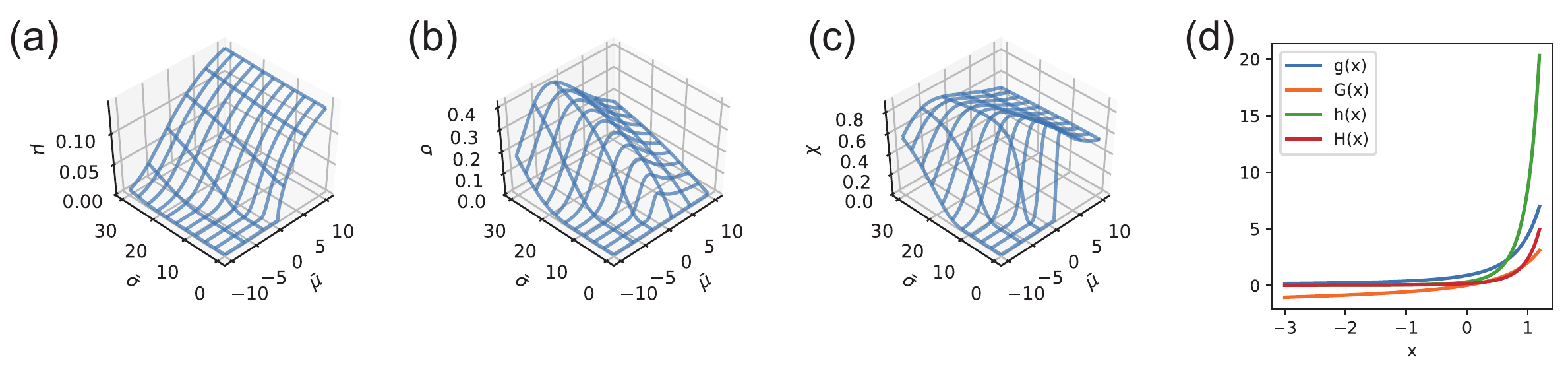}
\caption{The moment activation (MA). The MA maps the statistical moments $(\bar{\mu},\bar{\sigma})$ of the input synaptic current to those of the output spikes. The three components of MA are (a) the mean firing rate $\mu$, (b) the firing variability $\sigma$, and (c) the linear response coefficient $\chi$ used for the correlation map. (d) The family of Dawson-like functions appearing in the MA exhibiting super-exponential growth.
\label{fig:general} }
\end{figure*}

The three components of the MA, namely, the mean firing rate $\mu$ (Eq.~\ref{eq:mu}), the firing variability $\sigma$ (Eq.~\ref{eq:sigma}), and the linear response coefficient $\chi$ (Eq.~\ref{eq:linear_res}), are shown in Fig.~\ref{fig:general}(a)-(c). Each of these components depend on both the mean $\bar{\mu}$ as well as the standard deviation $\bar{\sigma}$ of the input synaptic current. The evaluation of the MA based on these integral representations becomes problematic both in reliability and in speed. First, the Dawson-like functions (Eq.~\ref{eq:g}-\ref{eq:H}) are ill-conditioned so that direct evaluation of these integrals may fail catastrophically. To illustrate this point, consider $g(x)$ of Eq.~\ref{eq:g}. When $x$ becomes increasingly negative, the exponential function outside the integral sign explodes whereas the exponential function inside the integral vanishes, resulting in `$\infty\cdot0$'-type numeric instability even for moderately negative values of $x$. Such scenario is frequently encountered in practice as negative values of $x$, corresponding to $\bar{\mu}>V_{\rm res}L$, happen to be in the biological range. The same kind of issue is further amplified in $h(x)$ since the integrand itself depends on $g(x)$. Second, even for the input range over which the functions are well-behaved, direct evaluation of the MA is slow as it involves double or triple integrals. In the following, we present an efficient numerical algorithm that overcomes these difficulties.

%% file: algorithms/tmp_recycled.tex
\section{Efficient numerical algorithm for the moment activation}

To achieve reliable and fast numerical evaluation of the MA for arbitrary input values, the overall strategy is to look for direct numerical approximations to the Dawson-like functions $g(x)$, $h(x)$, $G(x)$ and $H(x)$. This allows us to efficiently evaluate the inter-spike interval in Eq.~\ref{eq:mean_isi}-\ref{eq:var_isi} by computing $G(x)$ and $H(x)$ only at the integration bounds, thereby significantly reducing the computational complexity compared to explicit evaluation of the nested integrals. These approximations also enable efficient evaluation of the linear response coefficient $\chi$ (Eq.~\ref{eq:linear_res}) and the derivatives of the MA. Specifically, we seek approximations of the MA in four input current regimes, namely, the mean-dominant regime, the weak fluctuation regime, the subthreshold regime, and the intermediate regime. 

The mean-dominant regime corresponds to when the magnitude of input current mean is much larger than the input current standard deviation, that is, when $\vert\bar{\mu}\vert\gg \bar{\sigma}$. As a result, the Dawson-like functions (Eq.~\ref{eq:g}-\ref{eq:H}) explode or vanish, as shown in Fig.~\ref{fig:general}(d), rendering direct numerical integration intractable. To overcome this, we construct asymptotic expansions for each of the Dawson-like functions $g(x)$, $h(x)$, $G(x)$, and $H(x)$ with a suitable truncation. 
The weak fluctuation regime corresponds to when $\bar{\sigma}$ is close to zero regardless of the value of $\bar{\mu}$, in which case we derive explicit analytical expressions for the MA. The subthreshold regime corresponds to when both the input current mean and variance are small such that the output of MA vanishes. The intermediate regime corresponds to the input range outside the aforementioned three regimes. For this regime, Chebyshev polynomial approximations with look-up tables for the coefficients are used. In the following, we present details of these approximations for the MA under each input current regime. The derivatives of the MA are presented in Appendix.

%% file: algorithms/fast_dawson.tex
\subsection{The mean-dominant input current regime}

Here, we present asymptotic expansions of the Dawson-like functions (Eq.~\ref{eq:g}-\ref{eq:H}), which allow us to efficiently evaluate the MA when the input current mean is large, that is, when $\vert\bar{\mu}\vert\gg \bar{\sigma}$. 

The asymptotic expansion for $g(x)$ as $x\to -\infty$ is 
\begin{equation}
g(x)\sim \sum_{n=0}^\infty(-1)^{n+1}\dfrac{(2n-1)!!}{2^{n+1}x^{2n+1}}.
\end{equation}
For $x>0$, the following identity is used
\begin{equation}
g(x) = \sqrt{\pi}e^{x^2}-g(-x).
\label{eq:g_pos}
\end{equation}
In fact, the function $g(x)$ is related to the scaled complementary error function as $g(x) = \frac{\sqrt{\pi}}{2}{\rm erfcx}(-x)$, which has been implemented previously using a different approach\cite{Johsons_webpage}.

The asymptotic expansion for $G(x)$ as $x\to -\infty$ is
\begin{equation}
G(x)\sim -\frac{1}{4}\gamma_e  - \frac{1}{2}\log(-2x) -\frac{1}{8x^2}+\frac{3}{32x^4}-\frac{5}{32x^6} + \mathcal{O}(x^{-8}),
\label{eq:G_asym_neg}
\end{equation}
where $\gamma_e$ is Euler's constant. It is worth noting that $G(x)$ is well behaved for $x<0$ as the leading term in the asymptotic expansion is logarithmic. For $x>0$ we use the following identity
\begin{equation}
G(x) = \frac{\pi}{2}{\rm erfi}(x)+G(-x),
\label{eq:G_pos}
\end{equation}
where ${\rm erfi}(x)$ is the imaginary error function, a well known special function with existing numerical implementations. 

Using integration by parts, we obtain the asymptotic expansion for $h(x)$
\begin{equation}
h(x)\sim \sum_{n=0}^\infty \dfrac{a_n}{x^{2n+3}}.
\label{eq:h_asym_neg}
\end{equation}
Here, we supply the coefficient $a_n$ for the first few terms which are -1/8, 5/16, -1, 65/16, -2589/128, 30669/256, -52779/64. For $x>0$ we use the following identity
\begin{equation}
h(x) = \sqrt{\pi}e^{x^2}[\tfrac{1}{2}\log2+G(x)+G(-x)] - h(-x).
\label{eq:h_pos}
\end{equation}

Next, by integrating the asymptotic expansion of $h(x)$ term by term, we obtain the asymptotic expansion for $H(x)$ as 
\begin{equation}
H(x)\sim \sum_{n=0}^\infty \dfrac{-a_n}{2n+2}\dfrac{1}{x^{2n+2}},
\label{eq:H_asym_neg}
\end{equation}
where $a_n$ is the same coefficients in Eq.~\ref{eq:h_asym_neg}. No practically useful identity is found in the case of $x>0$ for $H(x)$. Therefore, we approximate $H(x)$ with the leading term in its asymptotic expansion as $x\to+\infty$
\begin{equation}
H(x)\sim \frac{\pi^2}{32}[{\rm erfi}(x)-1][{\rm erfc}(-x)]^2{\rm erfi}(x).
\end{equation}
Note that for numerical implementation, an appropriate level of truncation is applied to the asymptotic expansion to achieve a balance between accuracy and applicable input range.

The mean firing rate $\mu$ and firing variability $\sigma^2$ of the MA can then be evaluated by combining the approximations for $G(x)$ and $H(x)$ with Eq.~\ref{eq:mu}-\ref{eq:var_isi}. For the correlation mapping, we evaluate the linear response coefficient $\chi$ (Eq.~\ref{eq:linear_res}) using the derivative of the mean firing rate (see Eq.~\ref{eq:duu} in Appendix). 

%% file: algorithms/fast_maf.tex
\subsection{The weak fluctuation input current regime}

The weak fluctuation regime corresponds to when the input current variance $\bar{\sigma}^2$ is close to zero, regardless of the value of the input current mean $\bar{\mu}$. In this scenario, the integration bounds in Eq.~\ref{eq:mean_isi}-\ref{eq:var_isi} contain singularities as the input current variance $\bar{\sigma}\to 0$, making it unsuitable for numerical implementation. However, these singularities are removable as the moment activation is well behaved when $\bar{\sigma}\to 0$. We find that the corresponding limits exist and have vastly simplified analytical expressions as presented below.

The limit for the mean firing rate $\mu$ is
\begin{equation}
\lim_{\bar{\sigma}\to 0}~
\mu(\bar{\mu},\bar{\sigma})= 
\begin{cases}
0,& {\rm for }~~\bar{\mu}\leq V_{\rm th}L,\\
\dfrac{1}{T_{\rm ref} -\frac{1}{L}\log\left( 1-\frac{V_{\rm th}L}{\bar{\mu}} \right)},& {\rm for}~~\bar{\mu} > V_{\rm th}L.
\end{cases}
\label{eq:mu_sigma_zero}
\end{equation}
This limit is consistent with the solution of the leaky integrate-and-fire neuron model receiving a constant input current.

For the variance mapping, we note that the limit of the Fano factor as $\bar{\sigma}\to 0$ is simply the Heaviside step function
\begin{equation}
\lim_{\bar{\sigma}\to 0}
~\dfrac{\sigma^2}{\mu}=
\begin{cases}
1,& {\rm for }~~\bar{\mu}\leq V_{\rm th}L,\\
0,& {\rm for}~~\bar{\mu} > V_{\rm th}L.
\end{cases}
\label{eq:heaviside}
\end{equation}
Combining this result and Eq.~\ref{eq:mu_sigma_zero}, we conclude that
\[
\lim_{\bar{\sigma}\to 0}~\sigma (\bar{\mu},\bar{\sigma})= 0.
\]

For the linear response coefficient $\chi$ in Eq.~\ref{eq:linear_res}, the limit as $\bar{\sigma}\to 0$ is  
\begin{equation}
\lim_{\bar{\sigma}\to 0}
~\chi(\bar{\mu},\bar{\sigma}) = 
\begin{cases}
0,& {\rm for }~~\bar{\mu}\leq V_{\rm th}L,\\
\sqrt{\frac{2}{L}}\dfrac{1}{\sqrt{
T_{\rm ref} -\frac{1}{L}\log\left( 1-\frac{V_{\rm th}L}{\bar{\mu}} \right)}
\sqrt{\tfrac{2\bar{\mu}}{V_{\rm th}L}-1}
},& {\rm for}~~\bar{\mu} > V_{\rm th}L.
\end{cases}
\end{equation}
These limits can then be used to approximate the moment activation when $\bar{\sigma}$ is very close to zero, in which case Eq.~\ref{eq:mean_isi}-\ref{eq:var_isi} become numerically intractable. Similar limits can be derived for the gradient of the moment activation (see Appendix).

\subsection{The subthreshold input current regime}

The subthreshold regime corresponds to when both the input current mean $\bar{\mu}$ as well as the variance $\bar{\sigma}$ are weak so that the neuron receiving the input ceases firing. Concretely, this corresponds to when $I_{\rm ub}(\bar{\mu},\bar{\sigma}) = \tfrac{V_{\rm th}L-\bar{\mu}}{\sqrt{L}\bar{\sigma}}> C$ for some sufficiently large positive number $C$. In this scenario, the integrals in Eq.~\ref{eq:mean_isi} and Eq.~\ref{eq:var_isi} explode super-exponentially and all outputs of the moment activation including $\mu$, $\sigma$ and $\chi$ vanish. The quantity $C$ can thus be viewed as a form of generalized firing threshold.

\subsection{The intermediate input current regime}

The intermediate input regime corresponds to the input range outside the aforementioned regimes. In this regime, direct numerical integration for the Dawson-like functions is possible but slow. To overcome this, we follow the strategy previously used for implementing the scaled complementary error function by using Chebyshev polynomial approximations with look-up tables \cite{Johsons_webpage}. 

For $x \le 0$, we first apply the transformation $x'=\frac{4}{4-x}$, which maps the input $x\in(-\infty,0]$ to the unit interval $x'\in(0,1]$. We then subdivide the unit interval into $N$ subintervals of equal length, and fit (in the least-square sense) the function over each subinterval with a Chebyshev polynomial of an appropriate degree $K$. The coefficients of the polynomial expansion are then saved to a look-up table which can then be used for fast evaluation of each Dawson-like function. Special identities (Eq.~\ref{eq:g_pos}, Eq.~\ref{eq:G_pos}, and Eq.~\ref{eq:h_pos}) can then be used to evaluate the functions for $x>0$. This general strategy is applied to all of $g(x)$, $G(x)$, $h(x)$, $H(x)$ but with a couple of minor exceptions. First, for $G(x)$ subdivision is applied directly over the interval $x\in [-c,0]$ for some constant $c$ without the transformation because $G(x)$ does not vanish as $x\to -\infty$ and grows logarithmically to negative infinity. Second, since there is no special identity relating $H(x)$ with $H(-x)$, we have to apply Chebyshev polynomial approximation to $x>0$ as well, which is done by fitting $\tilde{H}(x) = H(x)e^{-2x^2}$ to a Chebyshev polynomia for each subintervals over $x\in (0,c]$.

%% file: algorithms/intnfire.tex
\section{Validating the MA with spiking neuron model}

As mentioned earlier, the MA (Eq.~\ref{eq:mu}-\ref{eq:linear_res}) is derived directly from the LIF neuron model (Eq.~\ref{eq:LIF}) through a series of mathematically rigorous approximations. There are three potential sources of error due to these approximations, namely, the diffusion approximation for the synaptic current, the assumption for stationary process, and linear response approximation for the correlation mapping. In this section, we quantify how well the MA approximates the LIF neuron model (Eq.~\ref{eq:LIF}) and numerically investigate the conditions under which these approximations are valid. 

The first potential source of error is the diffusion approximation which replaces the synaptic current $I_i(t)$ in (Eq.~\ref{eq:LIF}) representing input spikes with a Gaussian white noise with the same mean and variance. The mean and variance of the output spike train can then be derived by solving the first passage time problem associated with the firing threshold~\citep{Feng2006}. Under certain conditions, this input-output relationship of the LIF neuron model obtained from the diffusion approximation converges to the exact result.

We validate the diffusion approximation with a single integrate-and-fire neuron receiving a controlled synaptic current of the form
\[
I(t) = w_eS_e(t)- w_iS_i(t),
\]
where $S_{e,i}(t)=\sum_k\delta(t-t^k_{e,i})$ represent renewal processes describing the excitatory and inhibitory input spike trains and $w_{e,i}$ are the corresponding synaptic weights. We then control the statistical moments of the total input current $I(t)$ by fixing the synaptic weights to $w_e=0.1$ and $w_i=0.4$ while modifying the statistics of the input spike train $S_{e,i}$. We then simulate the LIF spiking neuron model (Eq.~\ref{eq:LIF}) under these settings and calculate the trial-averaged mean firing rate $\mu$ and firing variability $\sigma$ using a finite but large time window $\Delta t$ (see Eq.~\ref{eq:def_mean}-\ref{eq:def_cov}). According to the diffusion approximation, the output spike statistics of the MA should approach to that of the LIF neuron model when $w_{e,i}$ are sufficiently small and when $S_{e,i}(t)$ contain a sufficiently large number of spikes for a given period of time. 

\begin{figure*}[t]
\includegraphics[width=\textwidth]{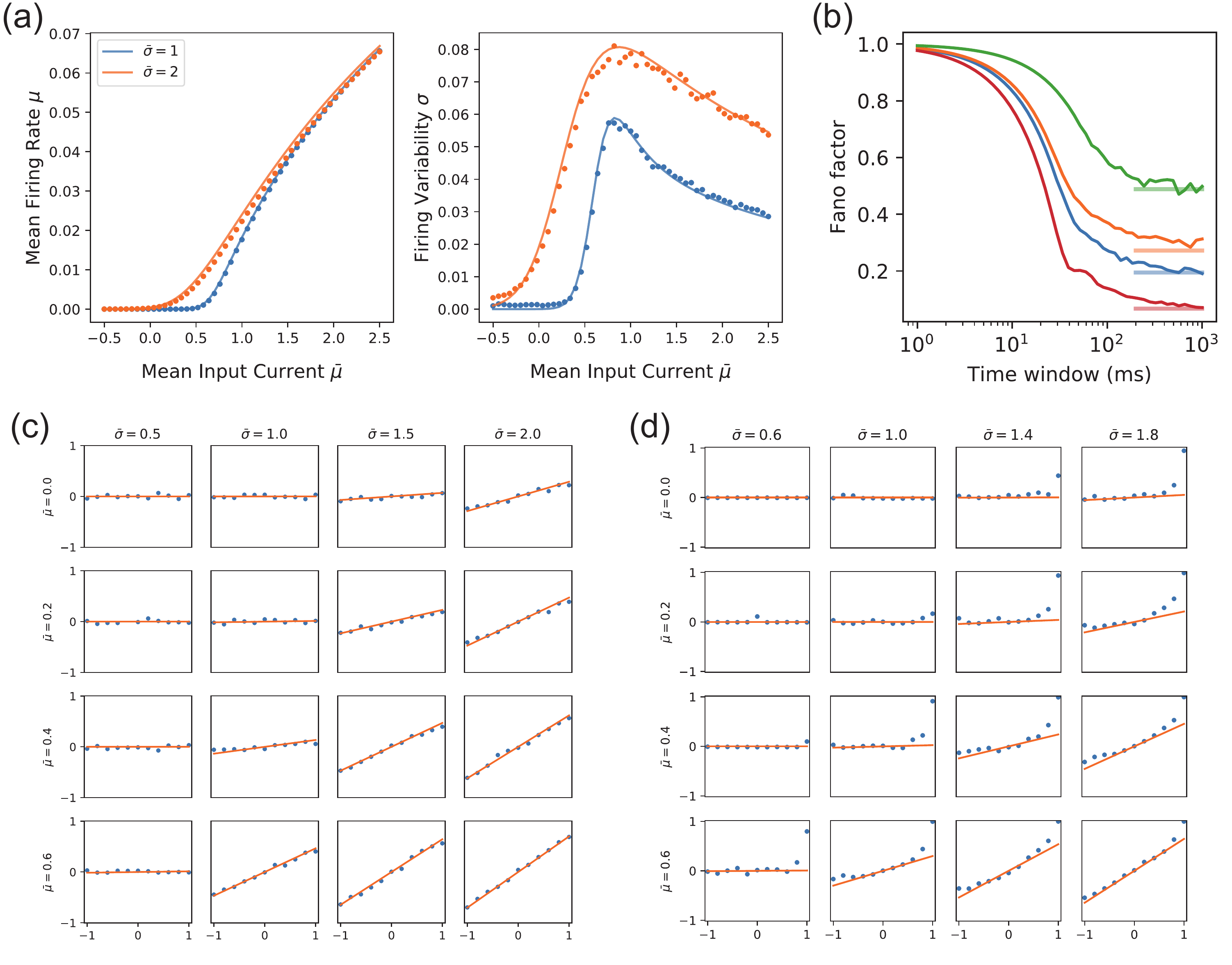}
\caption{Moment activation (MA) for approximating the firing statistics of spiking neuron model. (a) The mean firing rate $\mu$ and firing variability $\sigma$ as a function of the input current mean $\bar{\mu}$ for different input current variability $\bar{\sigma}$. The MA (solid line) show excellent agreement with the LIF neuron (dots) with the exception that the MA tends to underestimate the firing variability for inhibitory input currents. %\SI{}{\mV\per\ms}
(b) Empirical Fano factor (solid curves) decreases with the spike count time window and agrees reasonably well with the analytical predictions of MA (horizontal bars) for time windows larger than 200 ms. Each curve corresponds to a different neuron in the network. 
(c,d) Correlation map between a pair of LIF neurons receiving correlated synaptic currents with varying input conditions. Horizontal axis: input correlation; vertical axis: output correlation. In (c), we fix $\bar{\mu}_2=1$ and $\bar{\sigma}_2=1$ while varying $\bar{\mu}_1$ and $\bar{\sigma}_1$. The linear response theory provides accurate predictions (solid lines) to LIF neurons (dots). In (d), we vary both inputs with $\bar{\mu}_1=\bar{\mu}_2$ and $\bar{\sigma}_1=\bar{\sigma}_2$. The predictions based on linear response theory (solid lines) deviate away from LIF neurons (dots) for $|\bar{\rho}|>0$. This deviation becomes more apparent for inputs closer to $\bar{\rho}=1$.
\label{fig:IF_vs_maf}}
\end{figure*}

We visualize the results by plotting the mean firing rate $\mu$ and the firing variability $\sigma$ against the input current mean $\bar{\mu}$ at different values of input current variability $\bar{\sigma}$. As shown in Fig.~\ref{fig:IF_vs_maf}(a), the prediction of the MA (solid curves) as implemented using our algorithm agrees well with simulation results of the LIF neuron model (dots) for both $\mu$ and $\sigma$. Minor exceptions occur in the firing variability $\sigma$ in the subthreshold input current regime, in which the MA tends to underestimate the firing variability.

The second source of error may arise from the assumption for stationary process, a requirement for deriving the MA from the spiking neuron model. This assumption is reflected by the limit in the spike count time window $T$ in Eq.~\ref{eq:def_mean}-\ref{eq:def_cov}. This implies that the Fano factor, $\sigma^2/\mu$, computed by the MA corresponds to the infinite-time Fano factor. However, it is well known that the Fano factor of event count in a renewal process is time-dependent and that the Fano factor is one at $T=0$ and converges to a finite value as $T\to\infty$~\citep{Rajdl2020}. Fortunately, in practice the requirement for infinite $T$ can be relaxed to a finite but sufficiently large time window rather than an infinite one. To quantify how large is sufficient, we simulate a homogeneous recurrent network using the LIF neuron model with random synaptic weights and investigate the dependence of Fano factor on the size of spike count time window $T$. Specifically, the neural network consists of $N_e=100$ excitatory neurons and $N_i=100$ inhibitory neurons with a synaptic connection probability of $p=0.3$. The synaptic weights are drawn randomly from normal distributions such that $w_e\sim\mathcal{N}(0.2,0.1)$ and $w_i\sim\mathcal{N}(0.4,0.2)$. Each neuron in the neural network receives input currents in the form of Gaussian white noise with mean $\bar{\mu}_{\rm ext}$ and standard deviation $\bar{\sigma}_{\rm ext}$ which are drawn randomly for each neuron as $\bar{\mu}_{\rm ext}\sim \mathcal{N}(1,0.2)$ and $\bar{\sigma}_{\rm ext}\sim \mathcal{N}(1,0.2)$. 

As shown in Fig.~\ref{fig:IF_vs_maf}, the Fano factor (solid curve, each for a different neuron) computed from the spiking neuron simulation is equal to one for $T=0$ and gradually decreases as $T$ increases. After a sufficiently large time window, the Fano factor eventually converges to the theoretical limit (solid bars) predicted by the MA. We find that reasonably accurate approximation is achieved for time window larger than $T=200$ ms. This result also implies that the assumption for stationary process can be relaxed to weakly non-stationary processes, that is, processes with statistics that slowly change over a time scale much larger than $T=200$ ms.

The third source of error is the linear response theory used to obtain the pairwise correlation map of LIF neurons~\citep{LU2010913}. Conceptually, the linear response theory provides a linear-order approximation to the correlation map near $\rho_{ij}=0$, and is thus the most accurate for weakly correlated neural activity. However, as we show below through numerical simulations, this approximation is quite accurate for most input current values with some large discrepancy only for specific circumstances. Since we only concern pairwise correlations, it is sufficient to consider two neurons without loss of generality. We treat the aggregated post-synaptic currents as correlated Gaussian random variables without explicitly modeling the input spike trains. This also allows us to separate the effect due the linear response approximation from that due to the diffusion approximation. Mathematically, the input currents received by the pair of neurons are
\[
I_i(t) = \bar{\mu}_i + \bar{\sigma}_i \xi_i(t),
\]
where $\bar{\mu}_i$ and $\bar{\sigma}_i$ are the mean and standard deviation of the input current for neuron $i\in \{1,2\}$, and $\xi_i(t)$ are white Gaussian noise with correlation coefficient equal to $\bar{\rho}$. 
We compare the predicted spike correlation $\rho$ based on the linear response theory against the empirical spike correlation $\rho_{\Delta t}$ of the LIF neuron model for varying input current conditions. First, we fix $\bar{\mu}_2=1$ and $\bar{\sigma}_2=1$ while varying $\bar{\mu}_1$ and $\bar{\sigma}_1$. As shown in Fig.~\ref{fig:IF_vs_maf}(c), the linear response theory provides accurate predictions (solid lines) to LIF neurons (dots) even for correlation coefficients away from zero. Second, we vary both inputs with $\bar{\mu}_1=\bar{\mu}_2$ and $\bar{\sigma}_1=\bar{\sigma}_2$. As shown in Fig.~\ref{fig:IF_vs_maf}(d), the predictions based on linear response theory (solid lines) deviate away from LIF neurons (dots) for $|\bar{\rho}|>0$. This deviation becomes more apparent for inputs closer to $\bar{\rho}=1$. Based on these observations, it is evident that the MA based on linear response theory provides reasonably accurate predictions of the correlation mapping for a large region of input space and the quality of approximation only degrades when the moments of the input synaptic currents coincide.

%% file: algorithms/discus.tex
\section{Discussion}

%summary of the result
In this study, we have developed an efficient numerical algorithm for the moment neural activation (MA) through a combination of strategies that provide both reliability and speed. The proposed algorithm overcomes the numerical instability caused by a group of ill-conditioned integrals in the MA through asymptotic approximation, allowing reliable evaluation of the MA for arbitrary input range. Moreover, the proposed algorithm circumvents multiple nested integrals in the MA and reduces the computation to finite series expansion, thereby vastly reducing the computational cost for evaluating the MA. The proposed method is thus more effective than previous methods for evaluating neural firing statistics which require numerically solving the associated Fokker-Planck equation~\citep{Richardson07,Rosenbaum2016a}. The algorithm for evaluating the Dawson-like functions may also find application in other fields in which such ill-conditioned integrals occur.

%implication
The MA powered by the propose algorithm has several potential applications in neuroscience. %% correlated neural variability
Understanding the dynamics and functional roles of correlated neural fluctuations exhibited by neurons in the brain is a central question in neuroscience~\citep{Urai2022NatNeur}. Derived from spiking neuron models on a mathematically rigorous ground, the MA captures realistic correlated neural fluctuations of neural spikes and provides an ideal tool for tackling such a question. Specifically, the computational efficiency of the proposed algorithm may enable simulations of large-scale cortical circuits with correlated variability and provide new insights about cortical computation previously unobtainable with direct simulation of spiking neurons or simplified firing rate models. The efficient implementation of the derivatives of the MA also provides a tool for semi-analytical approach to investigating the dynamical properties of correlated neural fluctuations in neural circuits. By capturing the nonlinear coupling of correlated neural variability, the MA also emphasizes a stochastic view of neural spiking activity and can thus be applied to investigate probabilistic neural computation in the brain.

%limitations and future directions

The approach developed in this study may potentially be extended in two directions. First, the MA considered here is based on a particular type of spiking neuron model (the current-based LIF neuron model). In the future, the proposed method may be extended to other types of neuron models to incorporate biological features such as synaptic conductance and slow/fast synaptic time scales. Moreover, the MA considers pairwise neural covariance with zero temporal lag and future works may extend this to incorporating cross-covariance to fully capture the rich spatio-temporal covariance structure of cortical networks.

%% file: algorithms/derivatives.tex
\section*{Appendix: Derivatives of the moment activation}

In the following, we supply formulae for the derivatives of the MA. First, for the mean firing rate $\mu$, by differentiating Eq.~\ref{eq:mu} with respect to $\bar{\mu}$ and $\bar{\sigma}$ we obtain the corresponding partial derivatives
\begin{equation}
\dfrac{\partial \mu}{\partial \bar{\mu}}
=\frac{2}{L\sqrt{L}}\frac{\mu^2}{\bar{\sigma}}[g(I_{\rm ub})-g(I_{\rm lb})],
\label{eq:duu}
\end{equation}
and
\begin{equation}
\dfrac{\partial \mu}{\partial \bar{\sigma}}=
\frac{2}{L}\frac{\mu^2}{\bar{\sigma}}[g(I_{\rm ub})I_{\rm ub} - g(I_{\rm lb})I_{\rm lb}],
\end{equation}
respectively. Second, for the firing variability $\sigma$, by differentiating Eq.~\ref{eq:sigma} with respect to $\bar{\mu}$ and $\bar{\sigma}$ we obtain
\begin{equation}
\dfrac{\partial\sigma}{\partial\bar{\mu}}=\frac{3}{L\sqrt{L}}\dfrac{\sigma}{\bar{\sigma}}\mu[g(I_{\rm ub})-g(I_{\rm lb})]
-\frac{1}{2\sqrt{L}}\dfrac{\sigma}{\bar{\sigma}}\dfrac{h(I_{\rm ub})-h(I_{\rm lb})}{H(I_{\rm ub})-H(I_{\rm lb})},
\end{equation}
and
\begin{equation}
\dfrac{\partial\sigma}{\partial\bar{\sigma}}=\frac{3}{L}\dfrac{\sigma}{\bar{\sigma}}\mu[g(I_{\rm ub})I_{\rm ub}-g(I_{\rm lb})I_{\rm lb}]
-\frac{1}{2}\dfrac{\sigma}{\bar{\sigma}}\dfrac{h(I_{\rm ub})I_{\rm ub}-h(I_{\rm lb})I_{\rm lb}}{H(I_{\rm ub})-H(I_{\rm lb})},
\end{equation}
respectively. Third, for the linear response coefficient, the derivatives are
\begin{equation}
\dfrac{\partial \chi}{\partial \bar{\mu}}
= \frac{1}{2}\dfrac{\chi}{\mu}\dfrac{\partial\mu}{\partial\bar{\mu}}
-\frac{\sqrt{2}}{L}\sqrt{\frac{\mu}{\Delta H}}\left[I_{\rm ub}g(I_{\rm ub}) - I_{\rm lb}g(I_{\rm lb})\right]\frac{1}{\bar{\sigma}}
+\frac{1}{2\sqrt{L}}\chi\dfrac{\Delta h}{\Delta H}\frac{1}{\bar{\sigma}},
\end{equation}  
and
\begin{equation}
\dfrac{\partial \chi}{\partial \bar{\sigma}}
= \frac{1}{2}\dfrac{\chi}{\mu}\dfrac{\partial\mu}{\partial\bar{\sigma}}
-\dfrac{\chi}{\bar{\sigma}}\dfrac{2(I_{\rm ub})^2g(I_{\rm ub})-2(I_{\rm lb})^2g(I_{\rm lb}) +I_{\rm ub}-I_{\rm lb}}{\Delta g}
+\frac{1}{2}\dfrac{\chi}{\bar{\sigma}}\dfrac{I_{\rm ub}h(I_{\rm ub})-I_{\rm lb}h(I_{\rm lb})}{\Delta H},
\end{equation}  
where the short-hand notation $\Delta$ denotes the difference between a function evaluated at $I_{\rm ub}$ and $I_{\rm lb}$.

We also find analytical expressions of these derivatives in the weak fluctuation input current regime as $\bar{\sigma}\to 0$. First, for the mean firing rate $\mu$, by differentiating Eq.~\ref{eq:mu_sigma_zero} we obtain
\begin{equation}
\lim_{\bar{\sigma}\to 0}\dfrac{\partial\mu}{\partial\bar{\mu}} = 
\begin{cases}
0,& {\rm for }~~\bar{\mu}\leq V_{\rm th}L\\
\dfrac{V\mu^2}{\bar{\mu}(\bar{\mu}-VL)},& {\rm for}~~\bar{\mu} > V_{\rm th}L
\end{cases}
\end{equation}
For the derivative of $\mu$ with respect to $\bar{\sigma}$, the limit is found to be zero everywhere except an isolated singularity at $\bar{\mu}=V_{\rm th}L$. For practical purposes we simply set it to zero. 

Second, for the firing variability $\sigma$, the gradient with respect to $\bar{\mu}$ is zero at $\bar{\sigma}=0$ except that it is not well defined at $\bar{\mu}=V_{\rm th}L$. For numerical purposes we set it to zero, that is,
\begin{equation}
\left. \dfrac{\partial{\sigma}}{\partial\bar{\mu}}\displaystyle \right|_{\bar{\sigma}=0}
= 0,
\end{equation}
for all $\bar{\mu}$. The analytical limit for the derivative of $\sigma$ with respect to $\bar{\sigma}$ is
\begin{equation}
\lim_{\bar{\sigma}\to 0}
\dfrac{\partial{\sigma}}{\partial\bar{\sigma}}
= \frac{1}{\sqrt{2L}}\mu^{\frac{3}{2}}\sqrt{\frac{1}{(V_{\rm th}L-\bar{\mu})^2}-\frac{1}{\bar{\mu}^2}}.
\end{equation}

Third, for the linear response coefficient $\chi$, the limits of its derivatives are found to be
\begin{equation}
\lim_{\bar{\sigma}\to 0}
\dfrac{\partial\chi}{\partial \bar{\mu}}
=\frac{1}{\sqrt{2L}}\dfrac{1}{\sqrt{\mu(\frac{2}{V_{\rm th}L}\bar{\mu}-1)}}\dfrac{\partial\mu}{\partial \bar{\mu}}
-\sqrt{\frac{2}{L}}\frac{1}{V_{\rm th}L}\mu^\frac{1}{2}
\left(\frac{2}{V_{\rm th}L}\bar{\mu}-1\right)^{-\frac{3}{2}},
\end{equation}
for $\bar{\mu}>V_{\rm th}L$ and zero otherwise, and
\begin{equation}
\lim_{\bar{\sigma}\to 0}
\dfrac{\partial\chi}{\partial \bar{\sigma}} = 0.
\end{equation}

Note that in call cases, the derivatives vanish to zero for sufficiently large $I_{\rm ub}$.